\documentclass{aa}

\begin{document}
\newcommand{\eg}{{\sl e.g. }}
\newcommand{\ie}{{\sl i.e. }}
\newcommand{\etal}{{\rm et al. }}
\newcommand{\NaI}{$\rm Na~\sc I~$}
\newcommand{\CaII}{$\rm Ca~\sc II~$}
\newcommand{\chis}{$\chi^{2}~$}
\newcommand{\chir}{${\chi}_{\nu}^{2}$}
\newcommand{\kms}{$\,$km$\,$s$^{-1}$}

\newcommand{\rhounit}{\mbox{$M_\odot \,$pc$^{-3}$}}
\newcommand{\vunit}{\mbox{km\,s$^{-1}$}}
\newcommand{\Me}{\mbox{$M_\oplus$}}
\newcommand{\Msun}{\mbox{M$_{\odot}$}}
\newcommand{\Rsun}{\mbox{R$_{\odot}$}}
\newcommand{\Lsun}{\mbox{L$_{\odot}$}}
\newcommand{\Mdot}{\mbox{M$_{\odot}$~yr$^{-1}$}}
\newcommand{\lum}{erg~s$^{-1}$} 
\newcommand{\ltsimeq}{\raisebox{-0.6ex}{$\,\stackrel 
        {\raisebox{-.2ex}{$\textstyle <$}}{\sim}\,$}} 
\newcommand{\gtsimeq}{\raisebox{-0.6ex}{$\,\stackrel
        {\raisebox{-.2ex}{$\textstyle >$}}{\sim}\,$}} 
\newcommand{\prpsimeq}{\raisebox{-0.6ex}{$\,\stackrel
        {\raisebox{-.2ex}{$\textstyle \propto $}}{\sim}\,$}}

\newcommand{\UBV}{$U\!BV$}
\newcommand{\JHK}{$J\!H\!K$}
\newcommand{\JHKL}{$J\!H\!K\!L$}
\newcommand{\UBVRI}{$U\!BV\!RI$}
\newcommand{\AV}{\mbox{$A_V$}}

\title{Understanding the LMXB X2127+119 in M15}

\subtitle{I. X-ray eclipses and dips}

\author{Zach Ioannou\inst{1,}\thanks{Present address: Department of Astronomy, University of Texas at Austin, C-1400, Austin, TX 78712, USA},
        T. Naylor,
       \inst{1,2}
        A.P. Smale,
       \inst{3}
        P.A. Charles,
       \inst{4}
        K. Mukai
       \inst{3}
        }

\offprints{zac@astro.as.utexas.edu}

\institute{$^{1}$Department of Physics, Keele University, Keele, Staffordshire, ST5 5BG, UK\\
           $^{2}$School of Physics, University of Exeter, Stocker Road, Exeter, EX4 4QL, UK\\
           $^{3}$Goddard Space Flight Centre, NASA, Greenbelt, MD 20771, USA\\
           $^{4}$Department of Physics and Astronomy, University of Southampton, Southampton, SO17 1BJ, UK\\
           }

\date{Received}

\abstract{  
We present  X-ray observations of the  high-inclination low-mass X-ray
binary  system X2127+119  (AC211)  in the  globular  cluster M15  (NGC
7078). The observations consist of data acquired in 1996 with the {\it
RXTE} satellite and  in 1995 with the {\it  ASCA} satellite. Also, the
MPC1 data from the 1988  {\it GINGA} observations were de-archived and
re-analysed.\\
The  phase-folded 2-10  keV hardness  ratios from  all  three missions
differ significantly indicating that  the system can exhibit different
spectral behaviours.  We  find that the X-ray eclipse  profiles can be
described  relatively well using  a simple  model where  the secondary
star passes  in front of  a large X-ray  emitting region. For  this we
require a  mass ratio $(q=M_{1}/M_{2})$  of about one.  The  radius of
this  X-ray  emitting region  is  ${\sim}0.8R_{L1}$  and its  vertical
extent ${\sim}60^{\circ}$ above the orbital plane.  We suggest that if
this X-ray  emitting region were  an optically thick corona,  it would
explain various  puzzling aspects of  this system.  We also  show that
the X-ray dip observed at phases around 0.65 does not conform with the
idea that the dip is caused by vertically extended material associated
with  the stream/disc  impact  region, but  that  it could  be due  to
structure in the inner parts of the disc.
}

\authorrunning{Zach Ioannou et al.}
\titlerunning{Understanding the LMXB X2127+119 in M15}

\maketitle

\section{Introduction}

Low-mass X-ray  binaries (LMXBs) are interacting  binary star systems,
where the mass accreting star is either a neutron star or a black hole
with the mass-losing  secondary star $(M{\leq}1{\Msun})$ usually being
of late  spectral type.  Grindlay (\cite{Grindlay93}) lists  12 bright
LMXBs  that are  known to  exist  in globular  clusters.  The  general
properties of  these systems are summarised  in Table 1  of Deutsch et
al. (\cite{Deutsch00}).  The X-ray source X2127+119 is associated with
the globular cluster M15  (NGC 7078). The variable, ultraviolet-bright
star AC211, which lies very close to the core of M15, was suggested as
the  optical  counterpart by  Auri\'{e}re  et al.  (\cite{Auriere84}).
Charles  et al.  (\cite{Charles86})  spectroscopically confirmed  this
through the detection of a strong He{\sc ii} emission line.

X2127+119 belongs  to a  distinct group of  LMXBs. Namely,  LMXBs that
exhibit an  Accretion Disc  Corona (ADC). The  standard ADC  model was
first outlined by White et  al. (\cite{White82}) and McClintock et al.
(\cite{Mcclint82}) and it requires  a hot ionized cloud that surrounds
the accretion disc and which is supported by X-ray heating of the disc
by  a central  X-ray source.  The corona  could also  be  supported by
scattering  X-rays from  the central  source  onto the  disc and  thus
heating the disc material to temperatures of $10^{7}-10^{8}$~K.

Previous X-ray studies of X2127+119 with the EXOSAT satellite Callanan
et al.  (\cite{Callanan87}) revealed the presence of  periodic dips in
the    X-ray   light    curve    of   the    system.   Ilovaisky    et
al. (\cite{Ilovaisky93}) found that  the period of X2127+119 was 17.11
hours and that the system exhibited two very distinct dips both in the
X-rays and  in the  optical. Milgrom (\cite{Milgrom78})  first pointed
out that  systems possessing an  ADC would show orbital  variations in
their light  curve which are  due to both  the secondary star  and the
accretion   disc   only   if   their  inclination   is   high   enough
$(>75^{\circ})$.  In  this paper we present further  evidence that the
two dips  observed in X2127+119  are indeed due  to an eclipse  by the
companion star and by variations in the accretion disc structure. This
dip associated with  variation in the disc structure  is assumed to be
caused by the partial eclipse  of the central X-ray emitting region by
the bulge of  accretion disc material, which forms  at the point where
the  accretion stream from  the secondary  star impacts  the accretion
disc. However,  as we discuss  later in the  paper there is  a problem
with this ``standard'' interpretation concerning the cause of this dip
in X2127+119.

Homer   \&  Charles   (\cite{Homer98})  found   that   previous  X-ray
observations could also  be fitted with a parabolic  ephemeris as well
as     with     the    linear     ephemeris     of    Ilovaisky     et
al.    (\cite{Ilovaisky93}).    The    period   change    yielded    a
\.{P}/P~${\sim}~9{\times}10^{-7}$~yr$^{-1}$, which implies a very high
mass transfer rate of the order of ${\sim}6{\times}10^{-7}{\Mdot}$.

X2127+119 was  also found to  be an X-ray  burster by van  Paradijs et
al.  (\cite{vanpar90}), thus  firmly  establishing the  nature of  the
central  accreting object  as  a neutron  star,  which was  presumably
directly visible at that time.

In this  paper we  examine the complex  X-ray properties of  this high
inclination system. The nature of what  seems to be a very large X-ray
emitting region is  discussed, as well as the  argument of whether the
neutron star  at the center of  the system is normally  visible as the
observation of  bursts would  suggest. However, the  lack of  a strong
black-body component, the absence  of fast quasi periodic oscillations
and the  X-ray light  curve characteristics seem  to suggest  that the
neutron star, as well as the inner disc regions, are hidden from view.
Phase-resolved X-ray spectroscopy is performed using datasets from the
{\it  Rossi  X-Ray  Timing  Explorer  (RXTE)} and  the  {\it  Advanced
Satellite  for  Cosmology  and  Astrophysics (ASCA)}.   We  have  also
de-archived  and  re-analysed  the  MPC1  data from  the  {\it  GINGA}
satellite. Finally, in Sect. 5 we discuss various possible models that
could  account for  the complex  X-ray  behaviour of  X2127+119 and  a
summary of our conclusions is presented in Sect. 6.

\section{Observations and Data Reduction}

\subsection{The {\it RXTE} Data}

The {\it RXTE}  satellite observed X2127+119 twice in  1996. The first
observation (hereafter 96a)  was between 1996 April 3  15:08:08 UT and
1996 April 5  21:24:26 UT, which corresponds to  3.2 orbital cycles of
the system.  The second observation  (hereafter 96b) was  between 1996
November 26 05:50:24 UT and 1996 November 28 16:28:32 UT.

The  primary instrument  for these  observations was  the Proportional
Counter  Array (PCA),  which consists  of five  propane and  xenon gas
filled proportional counter units  (PCU). The energy response range of
the PCA instrument is 2-60 keV with an energy resolution of $<18\%$ at
6 keV.  The observations  for 96a were  carried out using  three units
(PCU0, PCU1 and  PCU2) with PCUs 3 and 4  switched off. In observation
96b all five PCUs were on but units 3 and 4 were periodically switched
on and  off. For  consistency we only  present here the  combined data
from PCUs 0, 1 and 2 for both observations.

Simultaneous observations  were also taken with the  High Energy X-ray
Timing  Experiment   (HEXTE).  The  HEXTE   instrument  comprises  two
clusters, each cluster composed of four phoswich detectors. The energy
response range for the HEXTE is  15-250 keV at an energy resolution of
15\% at 60 keV.

The PCA  data presented  here are the  {\it standard2a} mode  data and
were reduced with  {\sc FTOOLS} (see Giles et  al.~1995 and references
therein for  a full description of  the {\it RXTE}  spacecraft and its
experiments).  The data  were background  corrected but  not dead-time
corrected. The dead-time of the data  is less than 1\% for count rates
that are less  than 1000 counts/s per PCU. The  maximum count rate for
our  observations was about  220 counts/s  from the  sum of  all three
PCUs.

Data screening filters were  applied. These filters excluded data that
satisfied any  of the following criteria:  a) the time  since the last
passage  through the  South Atlantic  Anomaly (SAA)  was less  than 45
minutes  for observation 96a  and 30  minutes for  96b, b)  the source
elevation above  the Earth's limb  was less than $10^{\circ}$,  and c)
the spacecraft  pointing was offset more than  $0.02^{\circ}$ from the
source.

\begin{figure}
\vspace{7.5cm}          
\includegraphics{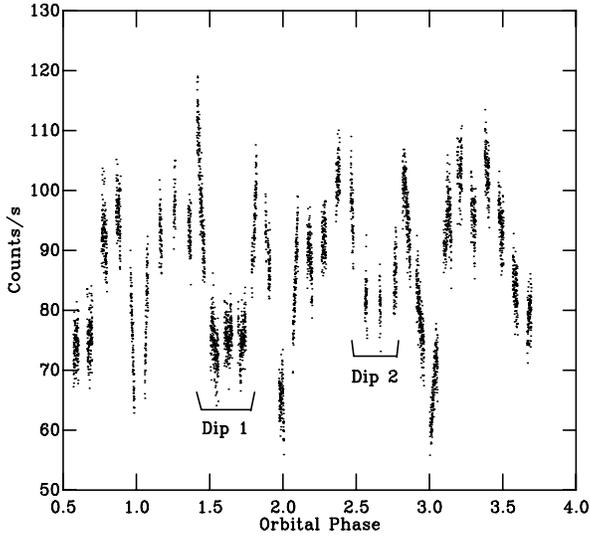}
\caption{2-15 keV X-ray light curve of X2127+119 from observation 96a
with the PCA instrument on {\it RXTE}. The binning is 16 seconds and
the phases were calculated using the ephemeris of Homer \& Charles (1998).}
\label{fig:pca2-15phase96a}
\end{figure}

\begin{figure}
\vspace{7.5cm}          
\includegraphics{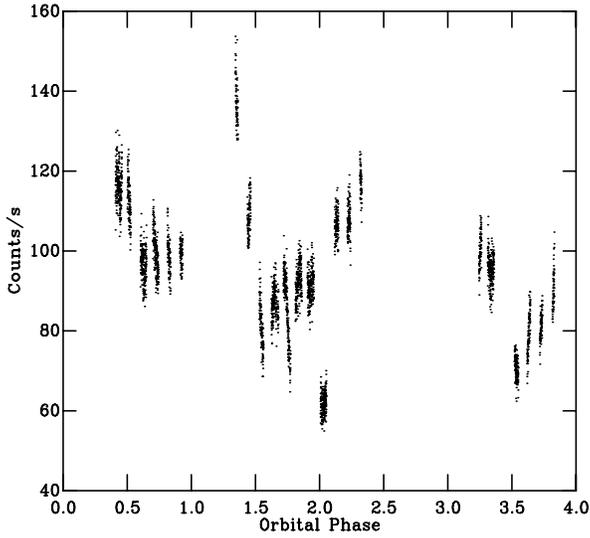}
\caption{2-15 keV X-ray light curve of X2127+119 from observation 96b 
with the PCA instrument on {\it RXTE}. The binning is the same as in 
Fig.~\ref{fig:pca2-15phase96a}}
\label{fig:pca2-15phase96b}
\end{figure}

This resulted in  67856 seconds of useful data,  arranged in 16s bins,
from  observation   96a  and  46560  seconds   from  observation  96b.
Figures~\ref{fig:pca2-15phase96a}  and  \ref{fig:pca2-15phase96b} show
the  2-15  keV   light  curves  for  the  96a   and  96b  observations
respectively.

HEXTE did not  detect X2127+119 in either the  96a or 96b observations
because  the background  count  rate completely  dominated the  source
signal. This is not surprising if  one looks at the way the background
affects the high  energy end of the PCA  spectra, where the background
starts to dominate the source signal at about 15 keV.

\subsection{The {\it ASCA} Data}

Observations with  the {\it ASCA}  satellite were carried  out between
1995 April 16 00:55:13 UT and 1995 April 17 03:50:15 UT, corresponding
to 1.54 orbital cycles.

Data selection criteria  were set such as to  exclude data whose times
included an  Earth elevation angle less  than $5^{\circ}$, geomagnetic
rigidity greater than 4 GeV/c, times of passage through the SAA, times
when the  pointing offset was greater than  $0.02^{\circ}$. Other high
background times  from the GIS instruments such  as contamination from
Earth's  radiation  belts  were  also  excluded.  The  data  were  not
background subtracted since the  background, as derived from blank sky
observations, always  constituted less than  0.5\% of the  source from
both the SIS and the GIS instruments.

The above screening requirements resulted in 35888 s of available data
from SIS  and 36432 s of data  from GIS, which were  arranged into 16s
bins.

The SIS data were reduced in {\it BRIGHT} mode and the X2127+119 count
rates in the 0.7-10.0 keV range were 8.9 counts/s and 7.3 counts/s for
SIS0 and SIS1 respectively. The corresponding count rates for GIS2 and
GIS3 were 5.5 counts/s and 6.3 counts/s respectively.

\begin{figure}
\vspace{7.5cm}        
\includegraphics{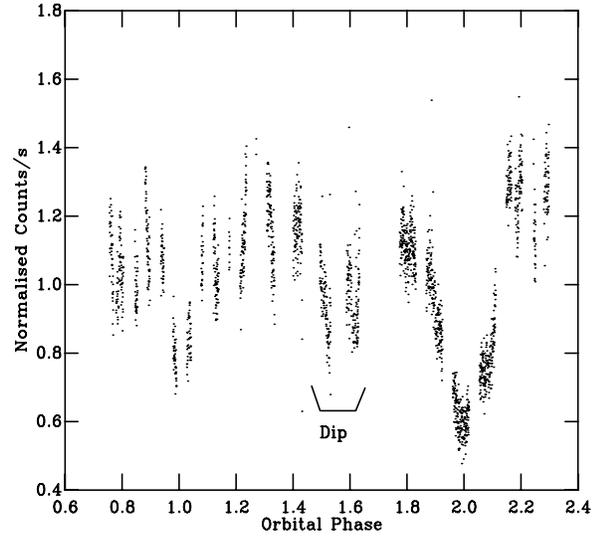}
\caption{Combined {\it ASCA} SIS0, SIS1, GIS2 and GIS3 0.7-10.0 keV average
light curve of X2127+119. The binning is 16 seconds.}
\label{fig:ascaall}
\end{figure}

\begin{figure}
\vspace{7.5cm}         
\includegraphics{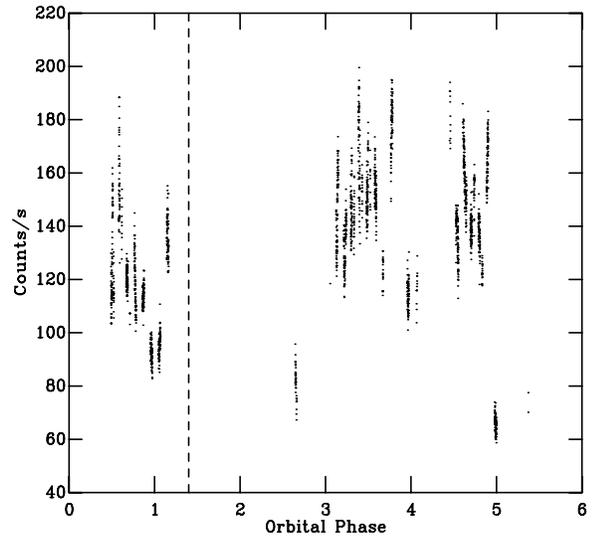}
\caption{The 2-18 keV light curve of the {\it GINGA} top 
layer MPC1 data. The dotted line represents the phase at which the 1988
X-ray burst occurred.}
\label{fig:gingafold}
\end{figure}

The combined SIS and GIS 0.7-10.0  keV average light curve can be seen
in  Fig.~\ref{fig:ascaall}. It  was created  by first  normalising the
data from the two SIS instruments by using the mean value of the light
curve count rate.  The mean count rate from SIS0 was  found to be 1.21
times higher than SIS1. Similarly  we normalised the data from the GIS
instruments and found that GIS3  contained 1.16 times more counts than
GIS2. The two resulting light curves from both SIS and GIS instruments
were then  normalised to  the mean SIS  level and finally  combined to
produce the  light curve shown in  Fig.~\ref{fig:ascaall}. The average
uncertainty resulting from combining the SIS and GIS data was found to
be at the 5\% level.

\subsection{The {\it GINGA} Data} 

The  {\it GINGA}  satellite observed  X2127+119 in  October  1988 (van
Paradijs  et  al.  1990)  using  the  Large  Area  Counter  (LAC).  We
de-archived the processed  and background subtracted standard products
from  the LEDAS  database and  have  re-analysed the  MPC1 mode  data.
Fig.~\ref{fig:gingafold} shows the 2-18 keV {\it GINGA} top layer MPC1
light curve.

\section{Temporal Analysis}

\subsection{The light curves}

\begin{figure*}
\begin{minipage}{12cm}
\includegraphics{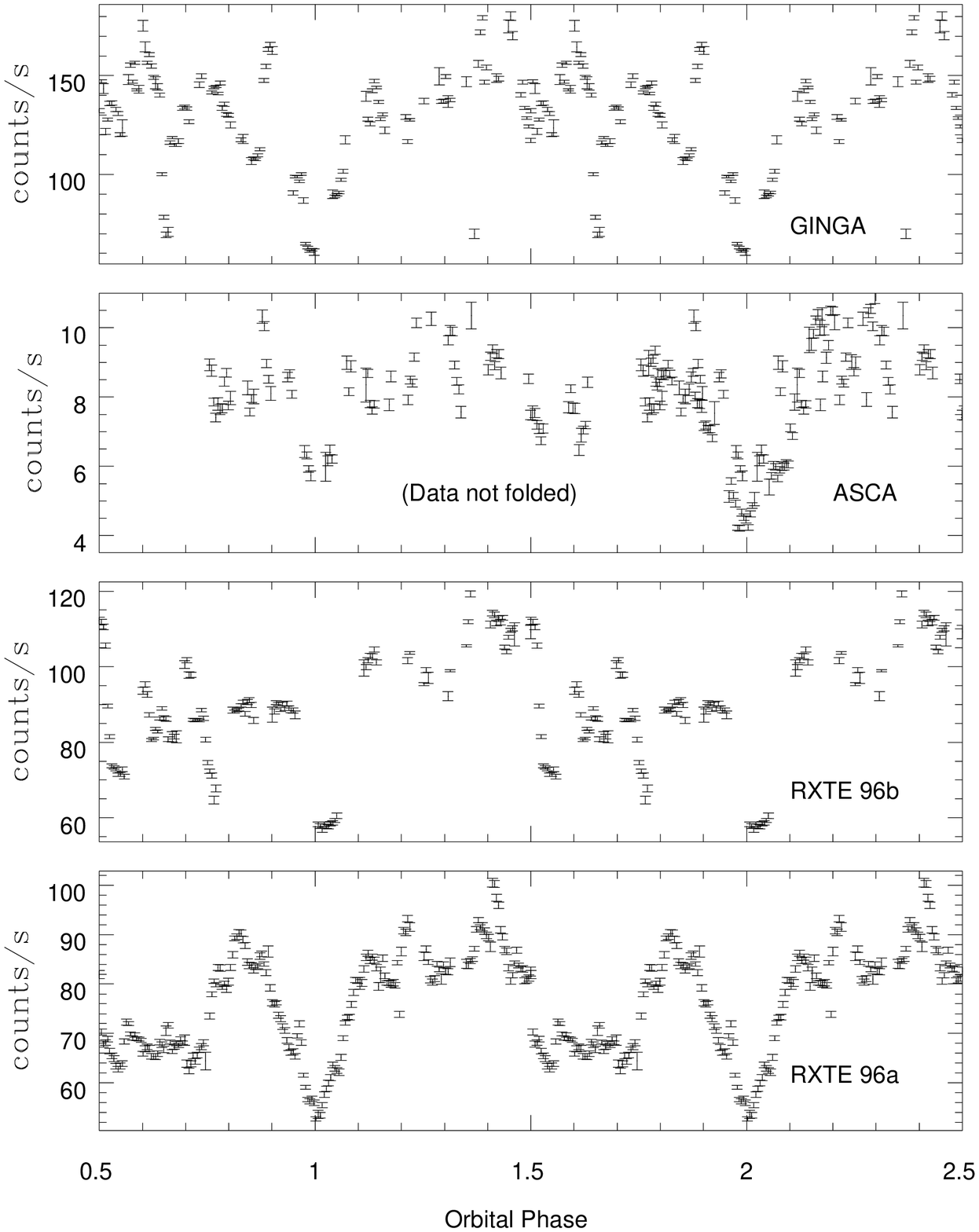}
\includegraphics{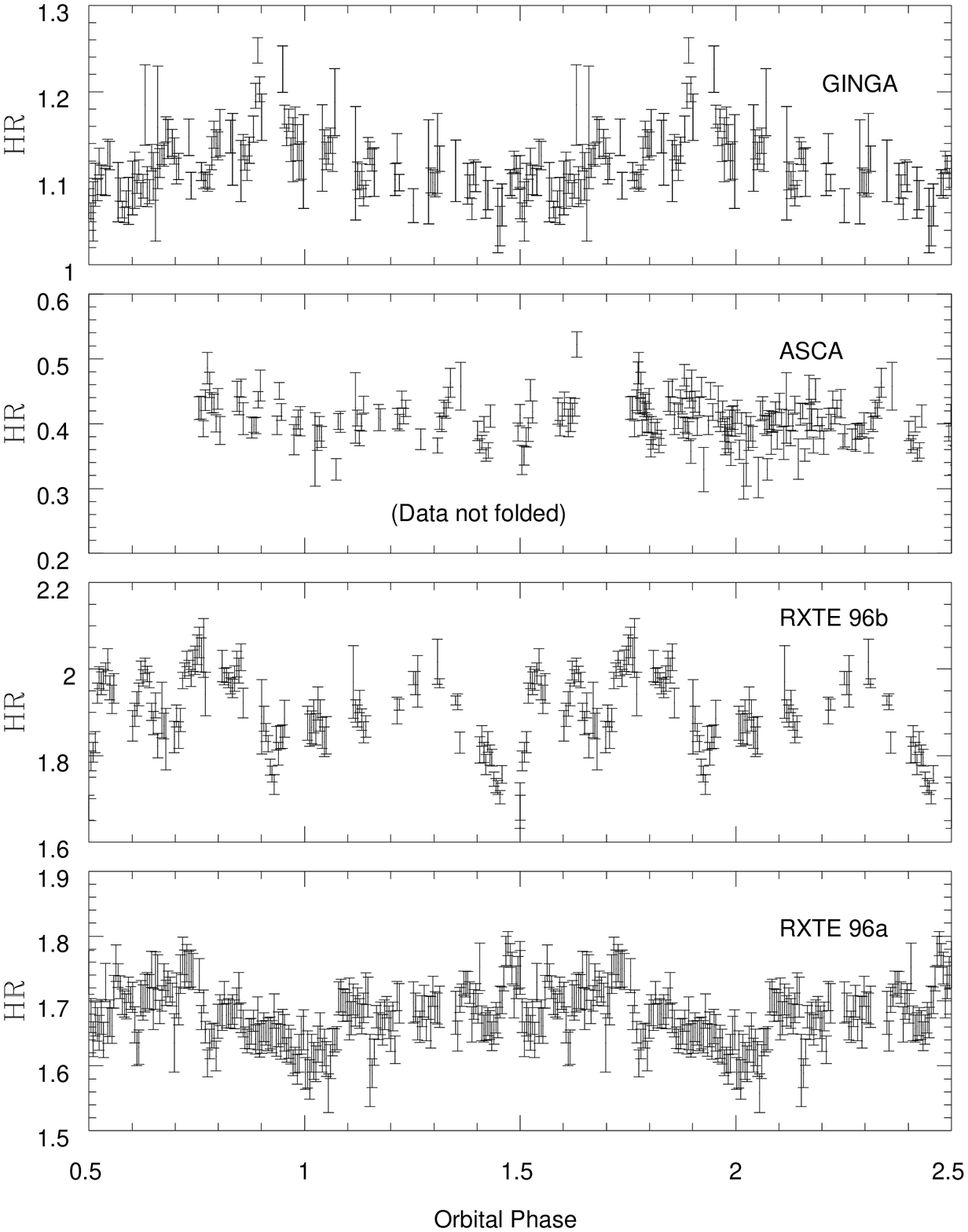}
\label{fig:lcshrs}
\vspace{12cm}
\end{minipage}
\caption{The phase folded 2-10 keV light curves (above left) and the
$(4-10)/(2-4)$ keV Hardness Ratios
(above right) are shown. All the data have a binning of $0.005$ in
phase or ${\sim}5$ mins. The {\sc ASCA} light curve and hardness
ratios have not been phase folded to illustrate the amount of change in the
eclipse profile in that dataset. Also, note the clear anti-correlation
of the HR plots between the {\sc RXTE} 96a and the {\sc GINGA} data.
The light curves and hardness ratios were folded on the ephemeris 
calculated by Homer \& Charles (1998).}
\end{figure*}

At the time  of our {\it RXTE} and {\it  ASCA} observations the linear
and the  quadratic ephemeris of Homer \&  Charles (\cite{Homer98}) are
aligned with  each other.  For this  reason we have opted  to fold the
light   curves  on   the  linear   ephemeris  of   Homer   \&  Charles
(\cite{Homer98}).  The  approximate  uncertainty  in phase  using  the
linear ephemeris  is about 0.001.  The  data from {\it  RXTE} show the
primary eclipse at phase 1.0 as  well as the dip, which occurs between
phases 0.5  and 0.8.   The {\it ASCA}  dataset covers  two consecutive
eclipses, and  it can clearly  be seen in  Fig.~\ref{fig:ascaall} that
the profile of the second  eclipse is substantially different from the
first. The dip that is extremely  prominent in the 96a {\it RXTE} data
is  not  very  well  defined   by  {\it  ASCA}.  In  the  {\it  GINGA}
observations the main  eclipse is deeper than with  {\it RXTE} and the
dip seems to be absent.

\subsection{Hardness ratios and Colour--Colour diagrams}

The $(4-10)/(2-4)$  keV hardness ratios  for all the  observations are
shown in  Fig. 5. Both hardness ratios,  from the PCA
data, show a  variation with orbital phase of about  10\% and there is
some  evidence  of correlation  with  X-ray  flux.  We find  that  the
hardness ratio  values drop slightly  during main eclipse.  During the
dip the hardness  ratio rises progressively as we  approach dip egress
and indeed  the hardness ratio plot  reaches its highest  level at the
point where the dip ends.

The {\it ASCA}  SIS (4-10)/(2-4) keV hardness ratio  against time plot
does not  show the same behaviour  as the {\it RXTE}  data. Whilst the
hardness  ratio  from  {\it  RXTE} exhibits  slight  modulations  with
orbital  phase,  what  we  observe   with  {\it  ASCA}  is  an  almost
featureless hardness ratio plot. The hardness ratio plot from the {\it
GINGA}  observations  (top right  of  Fig.  5)  exhibits a  completely
different behaviour  to the hardness  ratios from both the  {\it ASCA}
and the {\it RXTE} observations,  an indication that the system was in
a different  state than  the {\it ASCA}  and {\it  RXTE} observations.
The  hardness ratio is  evidently anti-correlated  with X-ray  flux as
originally reported by Callanan (\cite{Callanan93}).

\begin{figure}
\vspace{7.5cm}      
\includegraphics{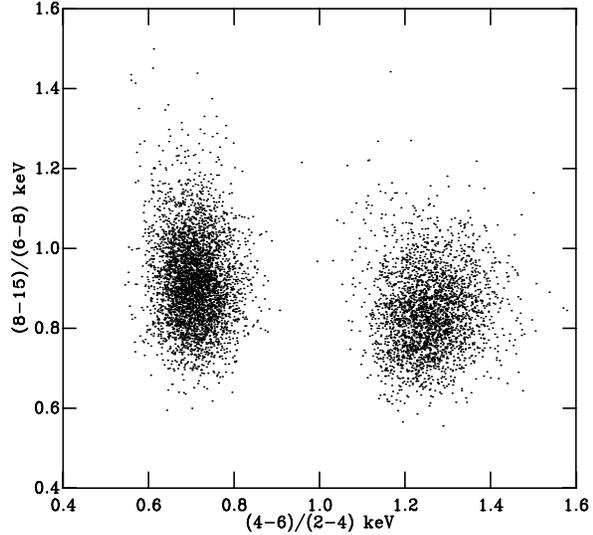}
\caption{Colour-colour diagram for {\it RXTE} data, corresponding 
to the count rate ratios for the (2-4),(4-6) keV and (6-8),(8-15) keV
bands respectively. The group of points on the left are the 
96a observation and the one on the right the 96b observation. The
offset between the two groups is real.}
\label{fig:colours}
\end{figure}

The colour-colour  diagrams constructed with the count  rates from the
(2-4),(4-6) keV  and (6-8),(8-15) keV bands  for both the  96a and 96b
{\it RXTE}  PCA data can  be seen in  Fig.~\ref{fig:colours}. Although
the data show  the same ``island'' behaviour as  found by van Paradijs
et al. (\cite{vanpar90}) they also clearly show that the system was in
a different state  in the two observations. The  ratio between the 4-6
keV and the 2-4 keV band is found  to be a factor of two higher in the
96b observation.

\subsection{The Main Eclipse}
 
Low mass  X-ray binaries that possess  an ADC and  are of sufficiently
high inclination  $(>75^{\circ})$ will exhibit both  eclipses and dips
in their light curves. This is  because the secondary star will now be
able to obscure part of the  ADC in addition to the obscuration caused
by the  accretion disc  rim structure. This  geometrical configuration
was first pointed out by Milgrom (\cite{Milgrom78}).

The   main  eclipses   in  X2127+119   are  broad   with   an  average
full-width-half-minimum   (FWHM)   of  about   0.15   of  an   orbital
cycle.  During the  main eclipse  the X-ray  flux from  the  system is
reduced  by about  50\%. This  is clearly  not an  eclipse of  a point
source (such  as the central compact  object) for two  reasons: a) the
eclipse profile does not show any sharp ingress/egress features, b) at
eclipse   minimum   the   profile   is   not   flat   as   one   would
expect. Furthermore,  at mid-eclipse the  flux does not drop  to zero,
indicating that there is also a significant contribution from extended
X-ray emitting regions that are not eclipsed by the companion star.

As there is no sharp ingress or egress from which to measure the exact
point of  the neutron star eclipse,  we can only make  guesses for the
inclination and  the mass  ratio of the  system. The  inclination must
obviously be high due to the presence of the eclipse and the dips.  We
therefore   begin  our   analysis  by   assuming  an   inclination  of
$90^{\circ}$. Considering the turn-off mass for the M15 cluster, Homer
\&  Charles  (\cite{Homer98}) assumed  a  mass  for  the secondary  of
$0.8{\Msun}$ and  the canonical value of $1.4{\Msun}$  for the neutron
star.  This yields a  mass ratio  value of  $q=M_{1}/M_{2}=1.75$. With
these basic assumptions  we can now attempt to  reproduce the observed
X-ray eclipses of the system.

For a $q$-value  of 1.75, a large X-ray emitting  region, of the order
of about $0.8 R_{L1}$ (where $R_{L1}$ is the distance from the neutron
star to  the inner  Lagrangian point) is  needed to match  the eclipse
width.

Lowering  the  mass  ratio  is  equivalent  to  introducing  a  larger
secondary star.  Including a larger  secondary star requires  a larger
X-ray emitting region  to keep the first and  the last eclipse contact
points at  the same phase.  Indeed, for lower  values of $q$,  such as
$q=1$, the X-ray emitting region requires a radius of $0.85 R_{L1}$ to
account  for the start  and end  contact points  of the  eclipse light
curve.

\begin{figure}
\vspace{7.5cm}        
\includegraphics{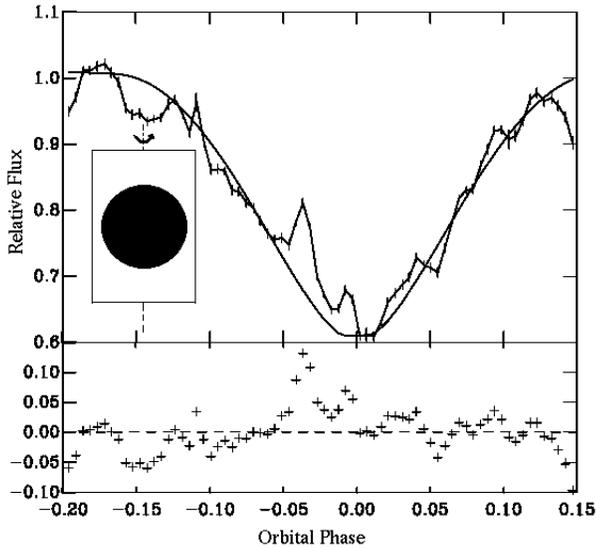}
\caption{A very simple model, which can reproduce the main eclipse
reasonably well is shown. The model (shown in the onset) is composed
of a dark disc (the secondary star) that passes in front of an X-ray 
emitting region (rectangle). The data is from the 96a {\it RXTE} 
observation.}
\label{fig:m15pic}
\end{figure}

For a  system with  an inclination of  $90^{\circ}$ and $q=1$  we find
that a  model consisting of  a black disc, representing  the secondary
star, which occults a rectangular X-ray emitting region, can reproduce
the eclipse light curve reasonably well. Figure~\ref{fig:m15pic} shows
the  model  light  curve  and  a  schematic  of  the  model  geometry.
Considering the  Roche-lobe geometry of  the system, the  aspect ratio
(i.e. height/radius)  of this X-ray  emitting region is 1.7,  with the
disc radius  having a value of  $0.85 R_{L1}$. This is  alarming as it
implies that  the height of the  X-ray emitting region  is larger than
its radius. We  have found that we could lower  the required height by
including a dark ``strip''  representing the accretion disc wall along
the orbital plane. The best preferred value for the full opening angle
of this  wall was  found to be  $25^{\circ}$. Including a  wall higher
than this  value had the effect  of flattening the  light curve during
mid-eclipse while lower  values for the strip required  an even higher
vertical  extent for  the X-ray  emitting  region to  account for  the
eclipse depth.

For values of $q>3$ we could  not find a disc radius and/or a vertical
extent for  the X-ray emitting  region that would produce  the eclipse
profile  observed.  The  model  light  curves were  always  wider  and
shallower than  the data as effectively  a $q>3$ results  in a smaller
secondary  star.  Thus  for  an  inclination of  $90^{\circ}$  we  can
constrain  $q$  to   $1<q<3$.  If  we  use  a   lower  inclination  of
$60^{\circ}$  then the  radius of  the X-ray  emitting  region becomes
smaller  at  about  $0.7  R_{L1}$  for  $q=1$  and  $0.6  R_{L1}$  for
$q=1.75$.  It  is therefore  evident  that  in  all of  the  geometric
configurations discussed  above, and even  for inclinations as  low as
$60^{\circ}$, we need a very  large X-ray emitting region with a large
vertical extent to account for the primary eclipse profile.

Although these  numbers must  be regarded as  highly uncertain,  it is
clear  that, to  explain  the eclipse  profile,  we need  to invoke  a
structure that  is relatively large  in comparison with  the secondary
star. An  ``accretion disk  corona'' (ADC) that  extends to  the outer
regions of  the accretion  disc and also  has a large  vertical extent
would thus  explain the observed  light curves. Scattered  X-rays from
the  ADC  could then  account  for the  non-zero  X-ray  flux when  at
mid-eclipse.

\begin{figure}
\vspace{6cm}        
\includegraphics{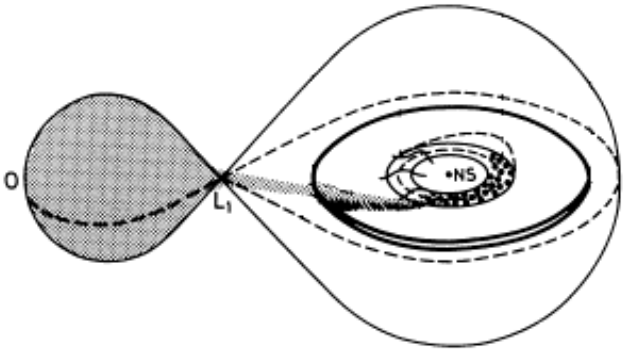}
\caption{Figure adopted from Frank et al (1987) showing the
interaction between the disc and the accretion stream. The accretion
stream penetrates the disc and forms a bulge at the circulirazation radius.}
\label{fig:streampic}
\end{figure}

\subsection{The Orbital Dips} 

Apart from the main eclipse  of the X-ray emitting regions, the system
also exhibits X-ray dips that are very prominent in the 96a {\it RXTE}
dataset.  The 96a  {\it RXTE}  coverage included  observations  of two
complete dips as  well as dip egresses and ingresses  at the start and
end of the  observation respectively. The coverage of  the dips in our
96b {\it  RXTE} observations  was not sufficient  to allow  a detailed
analysis and thus  our analysis of the dips depends  mostly on our 96a
observations.

The depth  of the dip  is about 70\%  of that of the  primary eclipse,
indicating that  the central  X-ray emitting regions  of the  disc are
eclipsed by some sort of disc structure. The drop in X-ray flux during
the dip is about 30\%. The  dip structure is seen to be persistent and
fairly  stable with  sharp  ingress  and egress.  We  observe the  dip
ingress and egress  in our {\it RXTE} data to have  a similar slope to
the ingress and  egress of the main eclipse. The  width of these X-ray
dips  is of the  same order  of the  primary eclipse.   This behaviour
indicates that the occulting bulge solid angle must be of the order of
the   secondary  star  and   also  possesses   a  very   well  defined
structure. We  have measured the depth  and width of  the two complete
dips observed in the 96a {\it RXTE} data, which are indicated as Dip 1
and Dip 2 in Fig.~\ref{fig:pca2-15phase96a}, and we find that they are
very dependent on energy (see Sect. 4.2).

The depth  of Dip 1  is seen to  remain roughly constant as  the X-ray
energy increases, but  its width becomes narrower. The  depth of Dip 2
is  seen to  decrease with  increasing  energy but  its width  remains
constant. As is also indicated  by the changing eclipse profile in the
{\it  ASCA} data, this  is evidence  that the  structure of  the X-ray
emitting region and/or of the accretion disc edge undergoes changes on
time scales of an orbital cycle.

\subsection{What is the cause of the dip?}

The full  width of the main eclipse  is 0.3 of an  orbital cycle.  The
full width of the dip is 0.4 of an orbital cycle. If the dip is caused
by  structure  at the  edge  of the  disc  then  this structure  spans
$145^{\circ}$ in azimuth.

It would be reasonable to assume  that the dip seen in the X-ray light
curves is caused  by the bulge of material created  at the point where
the accretion  stream impacts the accretion disc.   However, the phase
at which  the dip  ends, which  should be the  point where  the stream
first meets the disc, does not conform with what one would expect from
calculations of the ballistic trajectory of the accretion stream.

Using the  results of Lubow  \& Shu (\cite{Lubow75}) we  calculate the
ballistic trajectory of the stream and  we find that for a disc radius
of 0.8~$R_{L1}$ the stream impacts  the accretion disc at a point that
lies  about $25^{\circ}$  from the  line  of centres  between the  two
stars.  This translates to a  phase 0.93.  Phase 0.93 lies well within
the main eclipse, which extends back  to phase 0.8. The dip is seen to
end  at phase 0.75  which would  imply that  the point  of interaction
between the  stream and the disc  edge corresponds to  the point where
the line  of centres of  the two stars  and the line  representing the
radius of the disc are at $90^{\circ}$.

For the  ballistic trajectory of the  stream to intersect  the disc at
such  a  point,  the  disc  must  have a  very  small  size  of  about
0.1~$R_{L1}$. Clearly, such a disc  cannot give rise to the very broad
primary   X-ray   eclipse  observed.   Assuming   an  inclination   of
$90^{\circ}$ and  a mass ratio  of 1.7 we  find that to  reproduce the
main  eclipse width  and determine  the  contact phases  of the  X-ray
emitting region with the secondary star we need a disc that extends to
about 0.8~$R_{L1}$.

Frank  et al. (\cite{Frank87})  proposed a  model where  the accretion
stream from the  secondary star is shielded from  X-ray heating by the
accretion disc  and thus  is able  to penetrate the  disc down  to the
circularization  radius where  it shocks  with the  disc  material and
produces a bulge.  They show that the vertical extent of this bulge is
independent   of   system   parameters   and   that   it   is   always
${\sim}0.4R_{\rm circ}$,  where $R_{\rm circ}$  is the circularization
radius.    Using    a   $q=1.7$   the    circularization   radius   is
0.27~$R_{L1}$. At such  a radius the bulge would  occur at about phase
0.8 and  would subtend a full  opening angle of  about $20^{\circ}$ to
the central  regions of the  disc.  Our observations favour  the above
model and are not consistent with the bulge located at the edge of the
accretion disc.  Figure~\ref{fig:streampic}  adopted from Frank et al.
(\cite{Frank87}),  depicts   the  model  that  is   supported  by  our
observations of the orbital dips in X2127+119.
 
\section{Spectral Analysis}

\subsection{The Summed Spectra}

We found that multiple component models were needed to fit the overall
spectrum.  The  models fitted  included  {\it  power  law} (PL),  {\it
black-body}   (BB),  {\it  thermal   bremsstrahlung}  (BR)   and  {\it
Raymond-Smith}  (RS) hot  plasma emission  models, combined  with {\it
photoelectric absorption}  (wabs). We also  attempted to fit  the data
with  the Comptonisation  spectra of  Lamb \&  Sanford (\cite{Lamb79})
(CompLS)  and Sunyaev  \&  Titarchuk(\cite{Sunyaev80}) (CompST).   The
best  one component  model found  is a  photoelectrically  absorbed BR
model, which gave a {\chir} of 3.48 with 337 degrees of freedom (dof).
The best  two component model  found was a  photoelectrically absorbed
BB+BR model,  which returned  a {\chir} of  1.59 with 335  dof. Models
like  wabs(PL+CompLS) and  wabs(PL+BR) produced  fits with  {\chir} of
1.87 (334 dof) and 1.89 (335 dof) respectively.

Although, we  cannot draw any conclusions  from our model  fits to the
overall  summed spectra  we can,  by direct  comparison of  the summed
spectra in both  of the {\it RXTE} observations, see  that in the case
of the 96b  observation there is a surplus of soft  flux in the region
of 2-8 keV and that the spectra from both the 96a and 96b observations
are very similar in the 8-15 keV range.

\begin{figure*}
\begin{minipage}{75mm}
\includegraphics{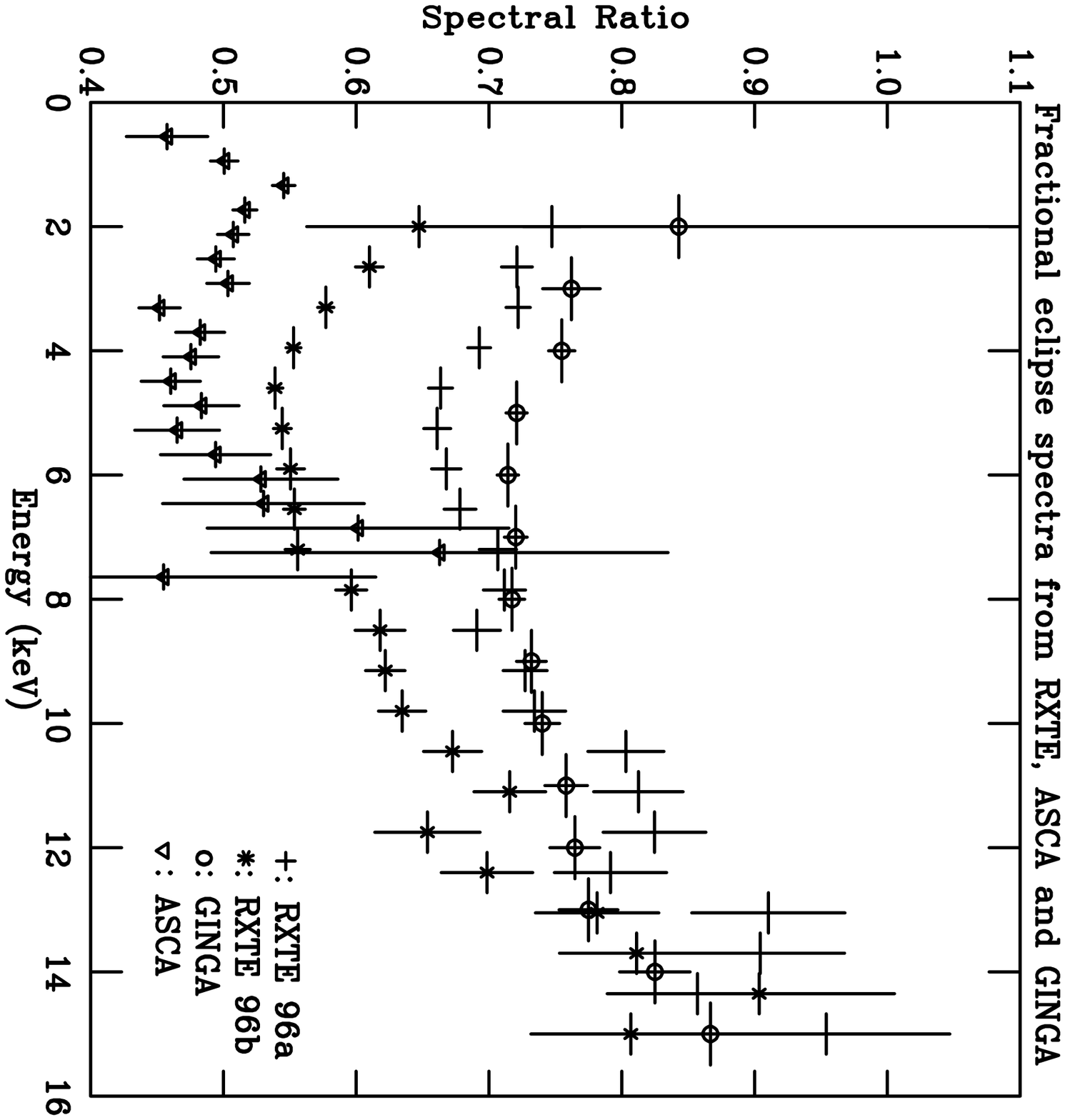}
\includegraphics{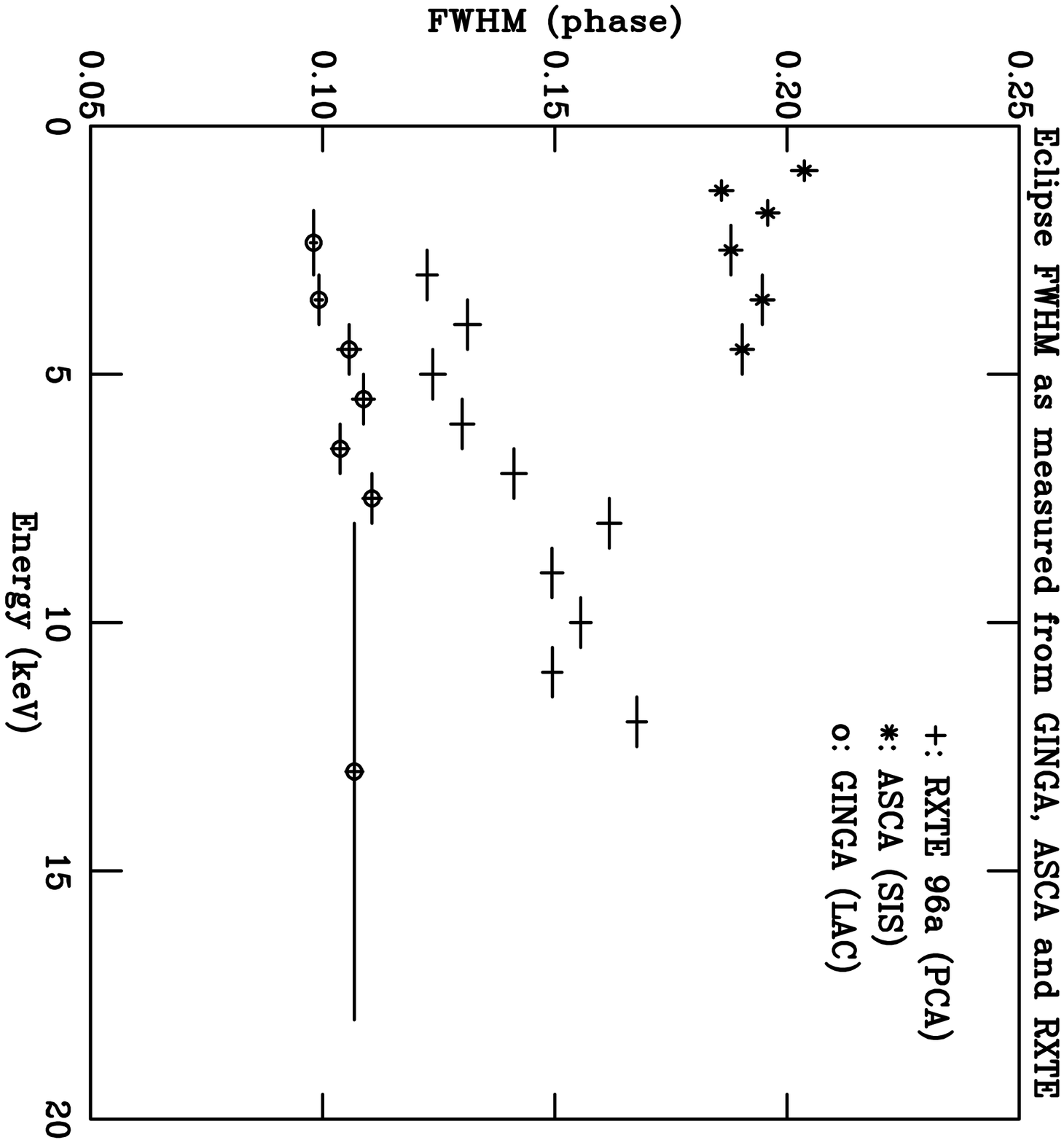}
\label{fig:eclpar}
\vspace{7cm}
\end{minipage}
\caption{Fractional eclipse spectra from {\it RXTE}, {\it ASCA} and GINGA 
(left) and eclipse FWHM (right). The main eclipse spectra were 
divided by spectra accumulated in the regions of 0.2-0.3 except in the 
case of {\it GINGA} where the phase coverage did not allow
that. Instead we used phases 0.4-0.6. The RXTE spectra are binned up by 
a factor of 2 and the {\it ASCA} data by a factor of 13 for clarity.}
\end{figure*}

\begin{figure*}
\begin{minipage}{75mm}
\includegraphics{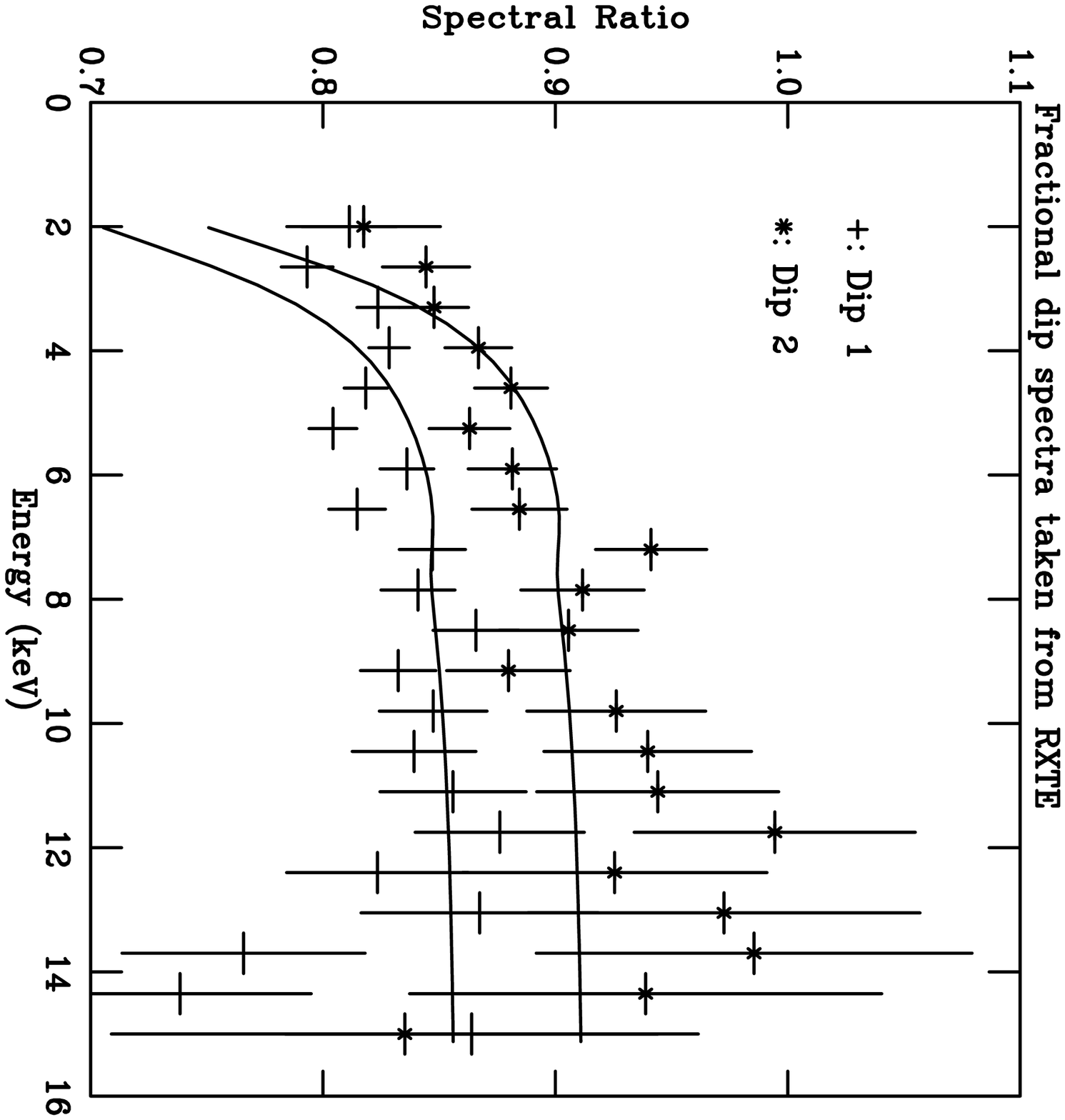}
\includegraphics{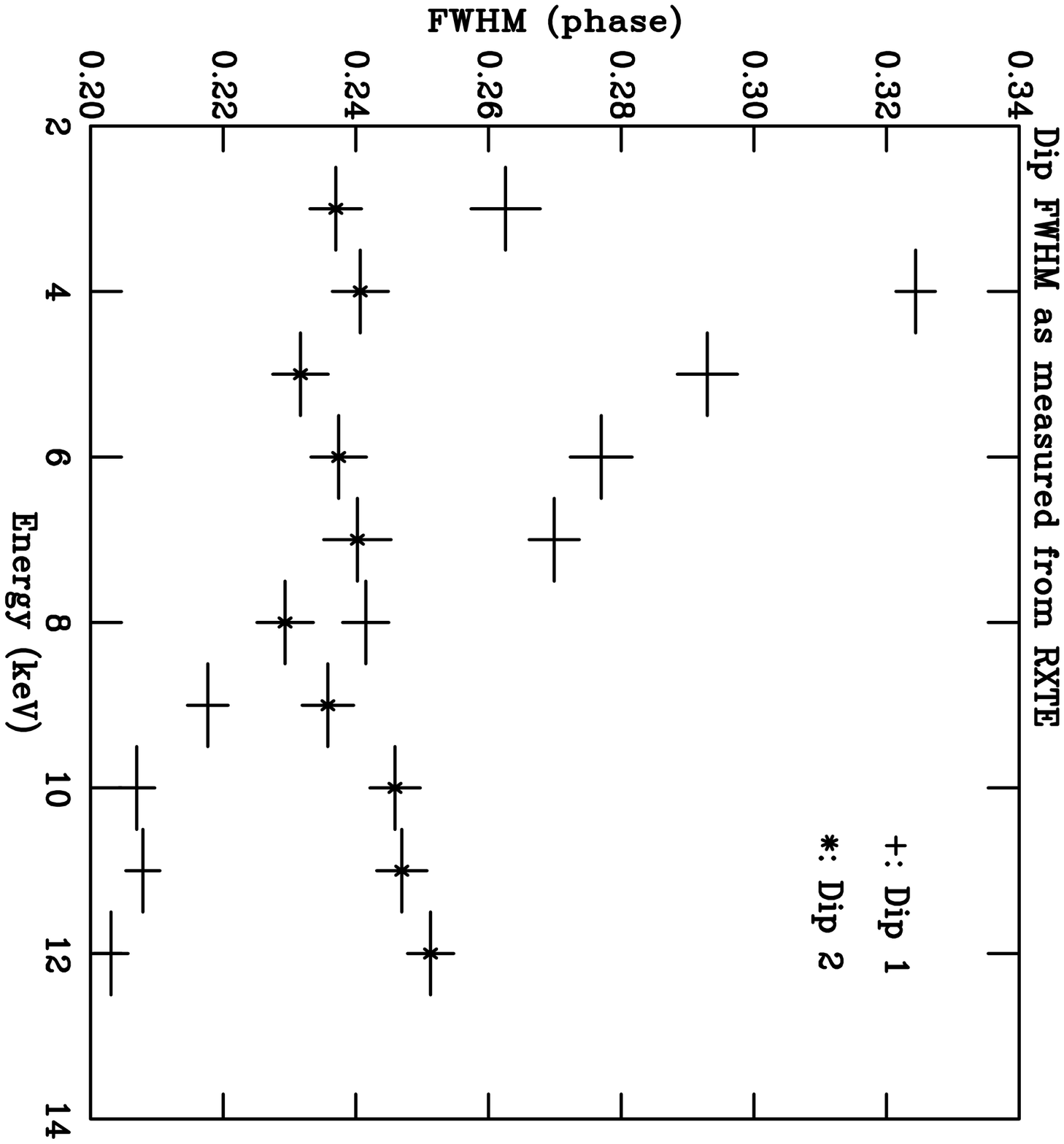}
\label{fig:dippar}
\vspace{7cm}
\end{minipage}
\caption{The fractional dip spectra (left) and dip FWHM (right) as measured
from the 96a {\it RXTE} observation. Dip 1 and Dip 2 correspond to the
dips as labeled in Fig.~\ref{fig:pca2-15phase96a}. The solid lines on
the fractional dip spectra plot shows the expected behaviour of photoelectric
absorption with energy.}
\end{figure*}

\subsection{Phase resolved spectroscopy}

In an attempt to try  and understand the physical processes occurring,
we examined spectra  from specific orbital phases.  We  split the {\it
ASCA} observations  into 11 orbital  phase regions from phase  0.12 to
phase 1.22.

It  was  found that  several  single  component  models were  able  to
reproduce the  spectra very  well. Photoelectrically absorbed  PL, BR,
CompLS  and Raymond-Smith  models  produced very  good fits  returning
{\chir} values  in the region of  1-1.5. It was  noticed however, with
the exception  of the absorption component, that  the model parameters
did not follow  any regular pattern and had  an erratic behaviour with
orbital phase. This leads us to the conclusion that the X-rays emitted
by this source are produced by complex processes and that these models
only provide smooth descriptions of the data.

Mitsuda et  al. (\cite{Mitsuda84}) proposed  a model where  the X-rays
observed in LMXB systems could  be explained by the superposition of a
single  temperature black-body  curve attributed  to the  neutron star
surface and  a multi-temperature  black-body component from  the inner
regions  of the  accretion disc.  Interestingly,  two-component models
that  included a  blackbody did  not produce  acceptable fits  for our
phase-resolved spectra ({\chir} never dropped below 2 despite the fact
that our  best model for  the summed spectra  was a BR+BB  model.  The
detection of a black-body component  in the X-ray spectra of X2127+119
would  imply that  we see  directly into  the innermost  parts  of the
accretion  disc  and  to  the  surface of  the  neutron  star  itself.
However, as  we do not detect  such a component we  believe that these
regions are obstructed from view by some kind of structure at the edge
of the disc.

Observations of X2127+119 by {\it BeppoSAX} and a brief examination of
the same {\it ASCA} spectra  presented here was performed by Sidoli et
al.  (\cite{Sidoli00}).   They found  that  they  could  fit the  {\it
BeppoSAX}  intensity averaged  spectra with  a partially  covered disc
black  body model  and a  power  law, where  the black  body flux  was
contributing between 24\% and 60\% of the total flux and that the same
model could be fit to the {\it ASCA} phase averaged spectrum. However,
as we have shown above the spectral properties of X2127+119 are highly
phase dependent and that the  overall averaged spectra can be modelled
relatively  well  by  a  variety  of models.  Furthermore,  a  partial
covering model  is not  able to explain  the behaviour of  the eclipse
depth with  energy that is seen  in the fractional  spectra of Fig.~9.
This further  emphasises the fact  that these models simply  provide a
smooth description of the data  rather than depict the actual physical
processes that take place in X2127+119.

Thus to study the spectral changes during the primary eclipses and the
dips  we  have  looked  at  the fractional  spectra  accumulated  from
different phases.  We divided the mid-eclipse and  the mid-orbital dip
spectra by  a spectrum accumulated  at phases (0.2-0.3).  The hardness
ratio at phases 0.2-0.3 remains constant indicating no abrupt spectral
changes.

We find  that the eclipse depth  is different in  each observation and
that it also  varies from eclipse to eclipse  but the overall spectral
behaviour is  stable over  the large time  period of  the observations
discussed here, with the  deepest eclipses produced at energies around
4-6 keV (see Fig. 9).

The fractional  spectra that correspond  to the middle of  the orbital
dips (phases 0.55-0.65) in our 96a {\it RXTE} observation suggest that
photoelectric absorption  could be  responsible for the  observed dips
and which is evident at  low energies (2-7 keV).  However, at energies
above 7 keV, where photoelectric  absorption is not expected to play a
dominant role in the spectra,  we find that the fractional spectra are
not equal to  unity (see Fig.~10). This could be  explained by a model
where a disc bulge whose structure is such that its dense central part
completely absorbs the X-ray flux from the centre of the disc and thus
lowering the  overall flux  observed, but its  less dense  outer parts
exhibit  normal  photoelectric absorption  behaviour  at low  energies
leading to the observed profiles.

What is  evident from the fractional  spectra of the  eclipses and the
dips   is  that   they  are   caused  by   two   completely  different
processes.  The fractional spectra  suggest that  the dips  are caused
partly   by  photoelectric  absorption,   while  during   eclipse  the
fractional  spectra are not  consistent with  photoelectric absorption
models.  Instead, they  show a  complicated profile  with  the deepest
eclipse of  the X-rays  around the 4-6  keV region. This  behaviour is
persistent in all of the  observations discussed here. If the eclipses
were  caused by  further  disc  structure one  would  expect that  the
fractional spectra would be similar to the spectra of the dip.

\subsection{Luminosity Considerations}

X2127+119 is known to exhibit variability in its long term X-ray light
curve.  Corbet  et  al.   (\cite{Corbet97})  reported  a  weak  37-day
modulation of the X-ray flux.

The 2-10  keV luminosity was  found to vary significantly  between the
four observations  reported in this  paper. Using the  widely accepted
distance of 10.5 kpc for M15  (Pryor \& Meylan 1993), we find that the
{\it  GINGA} MPC1  data from  1988  exhibit an  average luminosity  of
$3.2{\times}10^{36}$erg~s$^{-1}$.  The  {\it  ASCA} observations  from
1995  show  that  the  system   was  brighter  with  a  luminosity  of
$4.2{\times}10^{36}$erg~s$^{-1}$.  Also  the  {\it ASCA}  observations
show the highest  modulation of flux with a difference  of more than a
factor   of   two   between    the   minimum   luminosity   value   of
$2.3{\times}10^{36}$erg~s$^{-1}$    when    at    mid   eclipse    and
$5.0{\times}10^{36}$erg~s$^{-1}$  when at phase  0.22. The  system was
found to  be in a slightly  less bright state  at the time of  the 96a
{\it      RXTE}     observation      with     a      luminosity     of
$3.7{\times}10^{36}$erg~s$^{-1}$  and   again  brighter  in   the  96b
observation  with  a  luminosity of  $4.3{\times}10^{36}$erg~s$^{-1}$.
The slightly higher  luminosity of the 96b observation  as compared to
the 96a observation  combined with the excess soft  flux observed (see
Sect. 4.1)  suggests that the  system moved through  the colour-colour
space in a fashion that is typical for ``atoll'' sources (van der Klis
1997).

\subsection{Photoelectric absorption}

In our phase-resolved analysis of  the {\it ASCA} data (Sect. 4.2), we
found that various  one-component models were able to  fit the spectra
well. However,  it was noticed that  models such as CompLS,  BR and RS
preferred  roughly the  same amount  of photoelectric  absorption.  PL
models   generally   required  an   absorption   component  that   was
consistently ${\sim}$25\%  higher than other models.  The deviation of
the  $N_{H}$ values  between the  CompLS,  BR and  Raymond models  was
always less than 10\%.

Using  the  value  for  the  interstellar  reddening  towards  M15  of
$E_{(B-V)}=0.05$  (Pryor \&  Meylan 1993)  and the  empirical relation
between interstellar reddening and  X-ray absorption column from Ryter
et     al.     (\cite{Ryter75})     we     find    a     value     for
$N_{H}=3.4{\times}10^{20}$cm$^{-2}$.  This is a factor ${\sim}8$ lower
than  the  column  densities  that   are  required  to  fit  the  best
one-component   models   (CompLS)   to   our  phase   resolved   X-ray
spectra.  This suggests  that  a significant  amount of  photoelectric
absorption must be intrinsic to X2127+119.

\begin{figure}
\vspace{7.5cm}       
\includegraphics{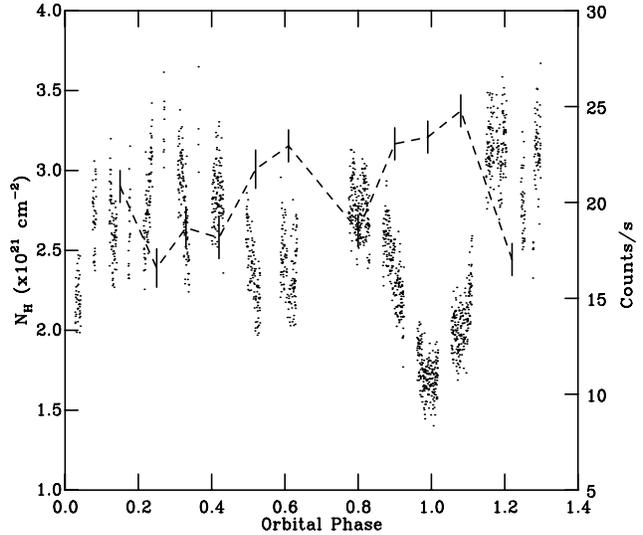}
\caption{Plot of the variation of $N_{H}$ with orbital phase as
observed with {\it ASCA}. The column densities were calculated using
the CompLS model. The 0.6-10.0 keV light curve is over-plotted
for comparison.}
\label{fig:nh}
\end{figure}

The modulation of  column density with orbital phase  in X2127+119 was
first    observed    by    Hertz   \&    Grindlay    (\cite{Hertz83}).
Fig.~\ref{fig:nh} shows a plot of the variation of column density with
orbital phase as derived from  our {\it ASCA} observation.  The column
densities were  calculated using  the CompLS model.  There is  a clear
anti-correlation between  column density  and X-ray flux.  The average
column  from all  the phase  resolved spectra  found using  the CompLS
model is $2.8{\times}10^{21}$cm$^{-2}$.  This is in agreement with the
$N_{H}$ values  quoted by Schulz  (\cite{Schulz99}) using observations
of X2127+119  from {\it ROSAT}.  However,  it should be  noted that in
all  of our  models, the  photoelectric  absorption is  assumed to  be
caused by material  with solar abundances and that  the column density
is  a  constant  factor  regardless  of  the  atomic  number  of  each
element. If the material of the system has a metallicity value similar
to that  of the M15 cluster  as a whole (i.e.  $[Fe/H]=-2.17$) then to
account  for all  the absorption,  that would  otherwise be  caused by
metals, we  would need to  multiply our column  values by a  factor of
${\sim}50$.  However, we  note that, despite the fact  that M15 has an
intrinsic  low metallicity there  is likely  to be  processed material
being transferred from the companion if it is evolved and convective.

Photoelectric absorption  could also be responsible  for the behaviour
of  the  hardness  ratio plots  from  the  {\it  RXTE} data  shown  in
Fig.~5. The rise  in the hardness ratio plot during  the dips could be
attributed  to  photoelectric absorption  of  the  soft  X-rays as  is
indicated by the fractional spectra of the dip shown in Fig.~9.

\section{Discussion}

White et al. (\cite{White81}) and McClintock et al. (\cite{Mcclint82})
were among the  first to suggest that the  broad eclipses, observed in
systems like X1822-371 and X2129+470, were due to the obstruction of a
large ionized cloud  of gas that surrounds the  accretion disc and the
central compact object.  Similar  ADC models were proposed by Callanan
et  al. (\cite{Callanan87}  and  Fabian et  al. (\cite{Fabian87})  for
X2127+119.   The   standard   ADC   model   is   reviewed   by   Mason
(\cite{Mason86}).

There are two controversial predictions from this model for X2127+119:
i) for high inclination systems the neutron star surface is never seen
directly, ii) as a result the source is X-ray faint because the system
is  viewed  through  an  optically  thin  scattering  corona.  In  the
following we  suggest that an optically thick  scattering corona could
account for the peculiar X-ray properties of X2127+119.

\subsection{High $L_{\rm opt}$ or low $L_{X}$?}

The low X-ray to optical luminosity ratio of $L_{X}/L_{\rm opt}=20$ is
normally given  as an  argument for X2127+119  being an ADC  source in
which  the observed  (but  not  intrinsic) $L_{X}$  is  low.  The  low
$L_{X}/L_{\rm opt}$  ratio is  taken to mean  that $L_{X}$ is  low. In
this section  we address the  question: {\it Is the  intrinsic $L_{X}$
really low?}.

Comparing  X2127+119 with  other  X-ray sources,  whose distances  are
relatively well  defined, namely  X1724-307 (Terzan 2),  X1746-370 and
X1850-087, we find  that the X-ray luminosity of  X2127+119 is similar
to the luminosities of the above  systems and that it is only a factor
of two fainter  than the galactic burst source  EXO 0748-676. The only
notable exception is the galactic  ADC source X1822-371 which is about
a factor of five fainter than X2127+119.

We would  like to draw attention  to the comparison  of X2127+119 with
the  source in the  globular cluster  Terzan 2,  which shows  no X-ray
eclipses  or  dips. Therefore  assuming  that  the  system is  of  low
inclination we can  expect that the X-ray flux  observed from Terzan 2
corresponds to the  bulk of the X-ray emission from  the centre of the
disc  and the  neutron star.  The  persistent 2-10  keV luminosity  of
Terzan  2 is ${\sim}6{\times}10^{36}${\lum}  (Guainazzi et  al. 1998),
which is very similar to that of X2127+119.

Substantial evidence therefore exists that points towards a relatively
normal $L_{X}$  level for  X2127+119, suggesting that  it is  the high
optical  luminosity  that is  responsible  for  the low  $L_{X}/L_{\rm
opt}$.  Indeed this  is evident if one looks at  the X-ray and optical
properties  of the  12 known  globular cluster  low mass  X-ray binary
systems that  are tabulated in Deutsch et  al. (\cite{Deutsch00}). One
can clearly see that although  the X-ray luminosity of X2127+119 has a
value  similar to  many of  the  other globular  cluster systems,  its
absolute optical magnitude  is by far the brightest.  In addition, the
very  broad  optical  eclipses  seen  in the  optical  counterpart  of
X2129+119  suggest that  the optical  emission also  emanates  from an
extended region of the system (i.e. the accretion disc and the ADC).

\subsection{Do we see the Neutron Star?}

Herein lies the heart of the X2127+119 puzzle.  The two main arguments
that point to the neutron star being hidden from view are: i) the lack
of a black-body  component in the phase- resolved  spectra in our {\it
RXTE} and {\it ASCA} datasets and ii) the X-ray eclipse profile, which
does not show evidence of a sharp ingress/egress.

In  contrast, results  from Dotani  et al.  (\cite{Dotani90})  and van
Paradijs et al. (\cite{vanpar90}), in their study of the burst seen in
1988, suggests that the  absence of evidence for reprocessed radiation
and the strength  of the burst ($>10,000$ counts/s  in the {\it GINGA}
LAC detector)  would imply that at  the time of the  burst the neutron
star surface  was directly  visible.  In addition,  the fact  that the
X-ray luminosity of this system  is comparable to the X-ray luminosity
of  presumably more face-on  systems (e.g.  Terzan 2)  and it  is also
relatively high  when compared to other  systems where we  know we can
see the  neutron star surface  directly (e.g. EXO  0748-676), suggests
that we do  in fact see the  full X-ray flux emitted. Yet,  how can we
have a high inclination ADC source which is X-ray bright?

\subsection{An optically thick corona?}

A very  large ADC  (or X-ray  emitting region) is  needed in  order to
account  for the  very  broad  eclipses and  dips  that are  observed.
Assuming a high  inclination system with $i=80^{\circ}-90^{\circ}$, an
ADC that extends out to  0.8~$R_{L1}$ is required to match the contact
points between the  limb of the secondary star  and the X-ray emitting
regions.   Such  a   large  ADC   is  not   unreasonable.   Fabian  et
al. (\cite{Fabian87})  pointed out that  in a low  metallicity system,
such as  X2127+119, a large  ADC is expected.  The reason is  that the
X-rays will stop  heating the surface of the disc  once they have been
sufficiently  absorbed.  In  a  low metallicity  system  photoelectric
absorption  is  not very  strong  and thus  the  X-rays  can heat  and
evaporate the surface  of the accretion disc out  to a larger distance
from the center. Using the  same model Fabian et al. (\cite{Fabian87})
were also  able to  produce a system  with high optical  luminosity by
heating  the accretion  disc  with scattered  X-ray  photons from  the
corona. This could therefore be the mechanism responsible for the very
high optical luminosity of X2127+119.

The  problem  with an  optically  thin corona  is  that  only a  small
fraction of  the X-rays produced at  the central source  of the system
would be  scattered into our line  of sight and thus  the system would
not exhibit its high X-ray luminosity when compared with other face-on
systems. Instead, we suggest that a large optically thick corona would
scatter the  centrally produced X-rays, diverting a  greater number of
X-ray photons into our line of sight producing a more nearly isotropic
radiation  pattern.  Hence  we  would  observe  most  of  the  emitted
luminosity.  In  addition,  the   two  reasons  stated  in  Sect.  5.2
concerning  the  visibility  of  the  neutron  star  itself  could  be
attributed to an optically thick corona, however, the main drawback of
this scenario is the lack of reprocessed radiation from the 1988 burst
(van Paradijs 1990).

\section{Conclusions}

From  our observations  acquired with  the {\it  RXTE} and  {\it ASCA}
satellites we find that:

i) the X-ray source X2127+119 must possess a very large accretion disc
corona. Assuming  an inclination of  $90^{\circ}$ we find that  we can
fit the X-ray  light curve with a simple model  consisting of an X-ray
emitting region whose vertical extent  is equal to or greater than its
radius.  The radius  of the ADC required is  about ${\sim}0.8 R_{L1}$.
This scenario  would require X2127+119  to have an  intrinsically high
mass transfer  rate in  order to support  such an  accretion disc/ADC.
Indeed,  X2127+119 is  the globular  cluster system  with  the longest
orbital period, which implies high mass transfer rates.

ii) an optically thick ADC could  account for the low X-ray to optical
luminosity ratio  and that it must  be the high  optical luminosity of
this system  which is responsible  for the low $L_{X}/L_{\rm  opt}$ of
20, not a low $L_{X}$.

iii) there  is strong evidence of photoelectric  absorption during the
X-ray  dips.  However,  spectra  taken  during the  main  eclipse  are
difficult  to  explain. They  do  require  a photoelectric  absorption
component  but they  also exhibit  various ``bumps''  and ``troughs'',
especially at low energies (2-4 keV), which we cannot account for. The
level of absorption is seen to be modulated with orbital phase.

iv)  models of  the  X-ray  spectra do  not  prefer strong  black-body
components, which  lead us to assume  that the surface  of the neutron
star is hidden from view, in  contrast with the {\it GINGA} results of
van  Paradijs  et  al.  (\cite{vanpar90})  who  did  not  observe  any
reprocessed radiation during the  large 1988 burst. The hardness ratio
plot  from the  {\it GINGA}  observation (which  differs significantly
from our  {\it RXTE}  and {\it ASCA}  observations) suggests  that the
system might  have been in a very  different state at the  time of the
burst.

v) an inconsistency exists with the position of the dip. A disc with a
radius  of $0.8  R_{L1}$ intersects  the ballistic  trajectory  of the
stream at an  angle of about $25^{\circ}$ from the  line of centres of
the two stars. However, the dip in  our 96a {\it RXTE} data is seen to
end at  about phase 0.75, which  implies an angle  of $90^{\circ}$.  A
situation where the stream penetrates the disc and produces a bulge at
the circularization radius could account for this effect.

\begin{acknowledgements}
We would  like to thank Dr.  Coel Hellier and Dr.  Edward Robinson for
useful discussions as  well as Dr. A. Beardmore for  his help with the
data reduction  phase. TN was a  PPARC Advanced Fellow  during most of
this work.  This research  has made use  of data obtained  through the
High  Energy  Astrophysics  Science  Archive  Research  Center  Online
Service, provided by the NASA/Goddard Space Flight Center and also the
Leicester Database  and Archive Service  at the Department  of Physics
and Astronomy, Leicester University, UK.
\end{acknowledgements}

\section{NOTE ADDED IN PROOF}

Following submission of this paper we have learned of the discovery by
the {\it  Chandra} X-ray observatory of  a second X-ray  source in the
core of M15.   These results are reported in  White \& Angelini (2001,
ApJ, 561,  L101). The new X-ray  source named M15 X-2  is actually 1.5
times  brighter  than  X2127+119.   The two  sources  were  completely
unresolved by all missions prior to {\it Chandra}.  However, since the
M15 X-2 shows very little or no variability, our analysis of the X-ray
light curve of the system  remains largely unaffected. The main effect
of  M15 X-2  on the  light curve  of X2127+119  is the  addition  of a
constant amount of  X-ray flux.  Thus we now  believe that the eclipse
of  X2127+119  is  actually  deeper  -- perhaps  even  total  --  than
previously thought,  further supporting the evidence  that this system
is of very high inclination,  perhaps very close to $90^{\circ}$. This
in turn would  have the effect of lowering the  vertical extent of the
ADC, since we no longer need to account for the extra X-ray flux.  The
discovery of M15 X-2 could  explain some of the peculiar properties of
X2127+119.   If the  X-ray bursts  originated from  M15 X-2  then that
explains our difficulty  in detecting the neutron star  in the spectra
of X2127+119. In addition, the strange behaviour of the hardness ratio
seen in Fig.~5  and our difficulties in modelling  the eclipse spectra
could also be attributed to contamination from M15 X-2.
 


\end{document}